\documentclass[manuscript]{acmart}
\usepackage{multirow}
\usepackage{hyperref}
\usepackage{amsfonts}
\usepackage{pifont}

\AtBeginDocument{%
  }

\copyrightyear{2023}
\acmYear{2023}
\setcopyright{acmlicensed}\acmConference[RecSys '23]{Seventeenth ACM Conference on Recommender Systems}{September 18--22, 2023}{Singapore, Singapore}
\acmBooktitle{Seventeenth ACM Conference on Recommender Systems (RecSys '23), September 18--22, 2023, Singapore, Singapore}
\acmPrice{15.00}
\acmDOI{10.1145/3604915.3609487}
\acmISBN{979-8-4007-0241-9/23/09}

\begin{document}

\title{The Effect of Third Party Implementations on Reproducibility}

\author{Bal\'{a}zs Hidasi}
\email{balazs.h@taboola.com}
\author{\'{A}d\'{a}m Tibor Czapp}
\email{adam-tibor.c@taboola.com}
\affiliation{%
  \institution{Gravity R\&D, a Taboola company}
  \streetaddress{Vill\'{a}nyi \'{u}t 40/B}
  \city{Budapest}
  \country{Hungary}
  \postcode{1113}
}

\renewcommand{\shortauthors}{Hidasi and Czapp}

\begin{abstract}
  Reproducibility of recommender systems research has come under scrutiny during recent years. Along with works focusing on repeating experiments with certain algorithms, the research community has also started discussing various aspects of evaluation and how these affect reproducibility. We add a novel angle to this discussion by examining how unofficial third-party implementations could benefit or hinder reproducibility. Besides giving a general overview, we thoroughly examine six third-party implementations of a popular recommender algorithm and compare them to the official version on five public datasets. In the light of our alarming findings we aim to draw the attention of the research community to this neglected aspect of reproducibility.
\end{abstract}




\maketitle

\newcommand{\cmark}{\ding{51}}
\newcommand{\xmark}{\ding{55}}

\section{Introduction} \label{sec:intro}
Reproducibility is the cornerstone of science as the scientific process relies heavily on the repeatability of experiments. Computer science is no exception to this rule. Experimentation in computer science is usually relatively cheap and quick that enables experiments to be executed relatively quickly and cheaply. As a negative side effect, quick and cheap experimentation can provide fertile ground for haphazard experiment design and careless experimentation. When a research field -- like machine learning -- is evolving rapidly and new algorithms are proposed every day, reproducibility tends to receive lower priority. This is then combatted by standardization efforts, e.g.~benchmark datasets, standardized evaluation frameworks, etc. Quickly evolving fields -- such as computer vision or natural language processing -- can still have their own benchmarks~\cite{deng2009imagenet,lin2014microsoft,kuznetsova2020open, rajpurkar2016squad, wang2019superglue, weston2015towards}. Compared to other areas of machine learning, reproducibility in recommender systems research is severely underdeveloped. This is probably connected to evaluation being one of the biggest challenges of the field. 

Recommendation quality can be assessed either online or offline. Online A/B tests provide better approximation on how well a recommender achieves its goals, but they are inherently not reproducible. The proprietary nature and commercial value of live recommender systems makes them accessible to only a very limited set of researchers, preventing the independent repetition of experiments. Due to the interactive nature of recommender systems, offline experimentation is just an imperfect proxy of the online setup. While reproducible in theory, multiple factors influence how repeatable an offline experiment is. In recent years, the research community has started discussing reproducibility and paying more attention to evaluation, including, but not limited to: 
\begin{itemize}
    \item \textbf{Public datasets.} Researchers are expected to showcase the performance of their algorithms on task appropriate public datasets. This is made possible by the increasing number of publicly released datasets, e.g.~\cite{requena2020shopper,ni2019justifying,kang2018self}
    \item \textbf{Open sourcing code.} More and more researchers release code along with their papers, which makes repeating the experiments significantly easier.~\cite{tang2018personalized,kang2018self,sun2019bert4rec}
    \item \textbf{Hyperparameter tuning.} Hyperparameters have significant effect on recommendation quality. Providing optimal hyperparameters for datasets and/or giving a general guide on how to get good parameterization is important for having strong baselines.~\cite{rendle2022revisiting,ferrari2021troubling,ferrari2020methodological}
    \item \textbf{Evaluation setups.} Data preprocessing, offline metrics, and executing appropriate experiments can make or break the evaluation part of a research project.~\cite{ferrari2021troubling,ferrari2020methodological}
    \item \textbf{Standardization, benchmarking.} Standardized benchmarks and frameworks supporting consistent evaluation over multiple algorithms allow for an informative comparison of models.~\cite{10.1145/3523227.3551472,argyriou2020microsoft,zhao2021recbole,said2014rival,sun2020we,zhu2022bars}
\end{itemize}

\subsection{Reimplementing recommender algorithms}
In this paper we discuss the reimplementation of recommender algorithms with respect to reproducibility. As far as we know, this is the first paper focusing on this topic. Reimplementing algorithms has been commonplace for recommender systems for decades. Before open sourcing code became widespread, researchers had to resort to reimplementing their baselines, based on the description alone. Even if an algorithm's code is publicly available, reimplementation might be reasonable due to \emph{convenience} and/or \emph{standardization}. Reimplementing an algorithm from scratch is also a good way of deeply understanding its inner workings.

\textbf{Convenience:} The official implementation might have an API unsuitable for the planned experiments. It might be written in a programming language the researcher is not familiar with. The researcher might need all baselines to be in a single (proprietary) framework for consistent evaluation. The official version might be implemented inefficiently or contain bugs. Etc.

\textbf{Standardization} in this case means reimplementing algorithms in benchmarking frameworks with the goal of providing fair, unified, and consistent evaluation over a large set of algorithms; instead of researchers using slightly different evaluation codes under the same name.

While unofficial reimplementations have existed for decades, open sourcing them is fairly new. When done correctly, this can be beneficial to the research community. Having multiple options can significantly increase accessibility. E.g.~researchers with less experience in programming might find it hard to utilize a C library in their experiments, but they might be able to use a Keras module. Releasing an unofficial implementation as part of benchmarking frameworks can contribute towards the standardization of evaluation. The authors of the original algorithm can also benefit from the wider visibility of their work and the contributors of the reimplementation can gain the goodwill of the community.

Unfortunately, there are also risks associated with unofficial releases. Reimplementing an algorithm from scratch can be a non-trivial task, especially if the algorithm is complex or has a lackluster description that lacks important nuances. The task is somewhat easier if a public official implementation exists since it can serve as a reference. Incorrect reimplementations -- that are not in line with the original -- can be harmful to research projects as they make baseline results invalid. Thus flawed, publicly available unofficial implementations can affect several research projects negatively.

Therefore, the quality and correctness of reimplementations is imperative, especially if they are publicly available. In this paper we investigate this topic. Since a thorough study of every unofficial implementation of every algorithm would be infeasible, we selected a single algorithm for this purpose. We chose GRU4Rec~\cite{hidasi2015session,hidasi2018recurrent}, a seminal work in session-based~/~sequential recommendations. GRU4Rec is an ideal candidate for our study for the reasons below:
\begin{enumerate}
    \item It is an older algorithm (released 7+ years ago), but has frequently been used as a baseline for session-based and sequential recommendations, due to it being a seminal work in this area.
    \item An official implementation\footnote{\url{https://github.com/hidasib/GRU4Rec}} was published along the papers, and it has been supported since then.
    \item The official version is implemented in Theano~\cite{al2016theano}, a now discontinued deep learning framework. While Theano still works fine, there is a clear incentive for reimplementation in more modern frameworks, e.g.~PyTorch.
    \item GRU4Rec is not a complex algorithm, but still has distinctive features that are crucial for its good performance.
    \item Papers report inconsistent results for GRU4Rec that might be a result of using incorrect implementations.~\cite{kang2018self,sun2019bert4rec,wu2019session,wang2020global,xu2019graph,luo2020collaborative}
\end{enumerate}

We aim to answer the following research questions:
\begin{itemize}
    \item (RQ1) Do third-party reimplementations implement the same algorithm as the official version?
    \item (RQ2) Are the third-party reimplementations feature complete?
    \item (RQ3) Are the third-party reimplementations correct? Are there any notable bugs or omissions?
    \item (RQ4) How do the differences between the reimplementations and the original affect offline measurements?
    \item (RQ5) How do the differences between the reimplementations and the original affect training times and scalability?
\end{itemize}

The rest of the paper is structured as follows. After a brief overview of GRU4Rec and its main features, Section~\ref{sec:compare} presents an in-depth qualitative comparison between six third-party reimplementations and the original (RQ1, RQ2, RQ3). Section~\ref{sec:quant} presents the results of our offline experiments (RQ4, RQ5). Our findings and their implications are summarized in Section~\ref{sec:discuss}. Section~\ref{sec:related} showcases related work, and Section~\ref{sec:conclusion} concludes this paper.
\section{Comparison of GRU4Rec implementations}\label{sec:compare}
In this section we compare the official GRU4Rec to six third-party reimplementations and answer RQ1, RQ2 and RQ3.

\subsection{Brief description of GRU4Rec}
GRU4Rec~\cite{hidasi2015session,hidasi2018recurrent} is a deep learning model originally proposed for session-based recommendations, and later utilized for other sequential recommendation tasks. The algorithm utilizes the Gated Recurrent Unit (GRU), a type of recurrent neural network. The model was released in 2015~\cite{hidasi2015session} and was improved upon in 2017~\cite{hidasi2018recurrent}. It revitalized the field of session-based~/~sequential recommendations and is considered to be one of the seminal works of the field.

The architecture of GRU4Rec is fairly simple. It receives an item of a session on its input and ranks every item based on how likely they are to be the next one. The inputted item is first embedded, then processed by one or more GRU layers. GRU layer(s) hold information on the state of the session, thus the model's inference is based on the whole sequence of events it observed so far in the session. Items are scored as the dot products between their embeddings and the hidden state. GRU4Rec has distinctive features adapting it to the specialties of recommender systems that distinguish it from simply processing behavior data with GRU. The official implementation contains all of the features including losses, embedding methods, negative sampling, etc., and it has many hyperparameters. The main features of GRU4Rec and the corresponding hyperparameters are listed below:
\begin{itemize}
    \item \textbf{Session-parallel mini-batches:} Since the length of sessions have a high variance and most of the sessions are short,~\cite{hidasi2015session} argues that (a) backpropagation through time (BPTT) is not necessary\footnote{It is hinted that a proprietary BPTT version exists for sequential tasks with longer sequences.}, and (b) events should be processed in a session-parallel way. A mini-batch is composed of input--target pairs from different sessions, each covering a single time step of its session. The next mini-batch is composed by shifting the input and the target to the next items respectively. When a session is fully  consumed, the next available training session fills its place, with its first and second items serving as the input and the target. The corresponding hidden state is also reset to zero. The size of the mini-batch is governed by the \texttt{batch\_size} parameter.
    \item \textbf{Negative sampling:} Scoring every item in every training step scales poorly, thus GRU4Rec utilizes two forms of negative item sampling.
    \begin{itemize}
        \item \textbf{Mini-batch based sampling:} Each example within a mini-batch uses the target items of the other examples as negative samples. This is a form of popularity based sampling that is very efficient to run on GPUs.
        \item \textbf{Extra shared sampling:} A number of \texttt{n\_sample} items are sampled for each mini-batch with sampling probabilities proportional to the items' relative frequency on the power of \texttt{sample\_alpha}. These negative samples are shared between the examples of a mini-batch.
    \end{itemize}
    \item \textbf{Losses:} Among the four available options (selected by \texttt{loss}), \cite{hidasi2018recurrent} deems two to be superior to the others.
    \begin{itemize}
        \item \textbf{Cross-entropy over softmax:} Item scores are passed through a softmax transformation\footnote{The final activation can be set to other than softmax, but it is not advised.}, before categorical cross-entropy loss is applied on them. LogQ correction~\cite{yi2019sampling} is available to offset the sampling bias affecting this loss specifically. Its strength is controlled by the \texttt{logq} parameter.
        \item \textbf{BPR-max:} Generalization of the BPR~\cite{rendle2012bpr} loss for multiple negative examples that fixes the vanishing gradient problem of using mean BPR. It also contains score regularization governed by \texttt{bpreg}. The final activation layer can be selected among linear, ReLU, ELU and sELU using the \texttt{final\_act} parameter.
    \end{itemize}
    \item \textbf{Embedding:} Items are embedded twice: for scoring, and as the input. There is one option for the former and three for the latter, which is governed by the \texttt{constrained\_embedding} and \texttt{embedding} parameters.
    \begin{itemize}
        \item \textbf{No embedding:} The input item is represented as a one-hot vector that is fed directly into the first GRU layer. For efficiency's sake, indexing is used instead of dense one-hot vectors and matrix multiplication.
        \item \textbf{Separate embedding:} A separate embedding is used for inputting and scoring the items.
        \item \textbf{Shared / constrained embedding:} The embedding on the input and the one used for scoring is the same.
    \end{itemize}
    \item \textbf{Training:} Multiple optimizers (AdaGrad, RMSprop, Adam, AdaDelta) are available for training (selected by \texttt{adapt}). The implementation defaults to Adagrad with Nesterov's momentum (\texttt{momentum} parameter). Learning rate is fixed during training and can be set by \texttt{learning\_rate}. The model is trained for \texttt{n\_epochs} number of epochs.
    \item \textbf{Regularization:} The algorithm uses dropout as the primary way of regularization, even though $\ell_2$ regularization is also available. The \texttt{dropout\_p\_hidden} parameter sets the drop probability for the GRU layers, and \texttt{dropout\_p\_embed} does the same on the input embedding. The latter has no effect in no embedding mode.
\end{itemize}
\subsection{Third-party reimplementations}
The selected set of third-party versions consists of the two Tensorflow and two PyTorch reimplementations, as well as two variants included in frameworks~/~algorithm collections.

\textbf{GRU4REC-pytorch}\footnote{\url{https://github.com/hungthanhpham94/GRU4REC-pytorch}, comparison is based on commit \texttt{666b84}} is a reimplementation in PyTorch. With 217 stars and 55 forks it is fairly popular. The first version was published in October 2018, and the last commit is from March 2021.

\textbf{Torch-GRU4Rec}\footnote{\url{https://github.com/yeganegi-reza/Torch-GRU4Rec}, comparison is based on commit \texttt{744f6b}} is another, less popular PyTorch reimplementation published in 2020.

\textbf{GRU4Rec\_Tensorflow}\footnote{\url{https://github.com/Songweiping/GRU4Rec_TensorFlow}, comparison is based on commit \texttt{d53fd9e}} is a relatively popular (213 stars, 89 forks) reimplementation in Tensorflow. Its updates were committed between March 2017 and June 2019.

\textbf{KerasGRU4Rec}\footnote{\url{https://github.com/paxcema/KerasGRU4Rec}, comparison is based on commit \texttt{239522}} is a Keras~/~Tensorflow reimplementation. The repository was forked from the official implementation in June 2018, but most of the code was rewritten before the Keras version was released in December 2018. The repository is still updated periodically by a bot, but the last meaningful update was committed in October 2020.

\textbf{Microsoft Recommenders}\footnote{\url{https://github.com/microsoft/recommenders}}~\cite{argyriou2020microsoft} is a popular collection of reimplementations of recommender algorithms, listed on the ACM RecSys github as one of the useful evaluation frameworks. According to its description, it contains examples and best practices for building recommender systems. It is often used for comparing algorithms. Its GRU4Rec version is reimplemented in Keras, and was last updated in October 2021. 

\textbf{Recpack}\footnote{\url{https://gitlab.com/recpack-maintainers/recpack}, comparison is based on commit \texttt{a83beb}}~\cite{10.1145/3523227.3551472} is an experimentation toolkit for top-N recommenders, listed on the ACM RecSys github as one of the useful evaluation frameworks. Its GRU4Rec is reimplemented in PyTorch.

\subsection{Comparison}

Five of the six third-party versions implement the same \emph{architecture} (RQ1) as the original. However, ``GRU4Rec'' of the Microsoft Recommenders collection only has two things in common with original: its name and that it uses GRU layers. The list of differences starts with the embedding layer, where this reimplementation utilizes extra information (item categories) besides the item ID. While changing the representation of the events within the session does not result in a new algorithm, it already undermines the goal of reproducing the original work. The list continues with the training of the GRU layer, where this algorithm requires sequences of equal length and processes multiple time steps in one forward pass, ignoring one of the distinctive features of the original. Scoring is also different and deeply flawed as well. In the reimplementation, a feedforward net scores hidden state--item pairs. This scoring function -- unlike that of the original -- can not be factorized into a sequence and an item dependent parts, thus the hidden state needs to be cloned as many times as the number of items to be ranked. This does not scale well beyond a few tens or maybe hundreds of items. Therefore, the number of negative item samples is severely limited, but more importantly, this algorithm can not rank items of even a small item catalog in a reasonable timeframe during inference. It seems that the authors are aware of these shortcomings, since the example showcasing this algorithm ranks only 50 items during evaluation, because full ranking would be impossible due to memory and time constraints. It is unclear why this completely different (and conceptually flawed) algorithm is called GRU4Rec in the Microsoft Recommenders framework that is supposed to be a curated collection of algorithms. Nevertheless, this reimplementation is excluded from further examination as it critically failed the first and most basic comparison.

Table~\ref{tab:features} showcases which reimplementation supports which \emph{features} of the original (RQ2). Surprisingly, no reimplementation supports shared embeddings, even though its usefulness on some datasets was highlighted in~\cite{hidasi2018recurrent}.

\begin{table*}
    \caption{Availability of GRU4Rec's main features in third-party versions: available (\cmark), missing (\xmark), or partially present~/~flawed (*).}
    \label{tab:features}
    \small
    \begin{tabular}{ll|ccccc}
      \toprule
      \multicolumn{2}{l|}{GRU4Rec feature} & GRU4REC-pytorch & Torch-GRU4Rec & GRU4Rec\_Tensorflow & KerasGRU4Rec & Recpack\\
      \midrule
      \multicolumn{2}{l|}{Session parallel mini-batches} & \cmark & \cmark & \cmark & \cmark & * \\
      \midrule
      \multirow{2}{*}{\shortstack[l]{Negative \\sampling}} & Mini-batch & \cmark & \cmark & \cmark & \xmark & * \\
      & Shared extra & \xmark & \cmark & \xmark & \xmark & * \\
      \midrule
      \multirow{2}{*}{Loss} & Cross-entropy & * & \cmark & \cmark & \cmark & \cmark \\
      & BPR-max & * & \cmark & \xmark & \xmark & \cmark \\
      \midrule
      \multirow{3}{*}{Embedding} & No embedding & \cmark & \cmark & \xmark & \cmark & \xmark \\
      & Separate & \cmark & \cmark & \cmark & \xmark & \cmark \\
      & Shared & \xmark & \xmark & \xmark & \xmark & \xmark \\
      \bottomrule
    \end{tabular}
    \end{table*}

Torch-GRU4Rec is the closest to the original as it supports all but the shared embedding feature. The other PyTorch implementation (GRU4REC-pytorch) has a limited feature set that is close to the base model from 2015, even though it is based on the improved model from 2017. The same is true for GRU4Rec\_Tensorflow, but with even more features missing. KerasGRU4Rec is the most basic one of the reimplementations, and it doesn't use negative item sampling. Instead, it computes the score for every item in every training step that makes it scale poorly. Recpack has two substantially different GRU4Rec implementations: the one with cross-entropy loss doesn't use negative sampling, the one with BPR-max does. Unfortunately, the implementation of the sampling is inefficient and also causes scalability problems (see Subsection~\ref{ssec:scaling}). Recpack also does not replicate session-parallel mini-batches correctly, as each of its batches lasts until every session within it is consumed. This increases training time, and also affects the loss.

Besides the hyperparameters corresponding to missing features, reimplementations might lack other ones as well. E.g., momentum with Adagrad and logQ correction is not available in any of the third-party versions. The former is missing due to the built-in Adagrad optimizers of PyTorch and Tensorflow lacking that feature. In some versions, dropout -- especially embedding dropout -- is also missing (or does not work). Hyperparameters are hard-coded in certain reimplementations, with KerasGRU4Rec being the most extreme example with hard-coded layer size and dropout probability.

Torch-GRU4Rec is mostly error free, but an easily noticeable typo prevents it from running as is. Other reimplementations contain a fair amount of \emph{bugs and errors} (RQ3). Some bugs are relatively easy to notice, while others are more insidious. The effort required for fixing the bugs varies widely. The nature of these bugs can be (a) incorrect implementation, e.g.~flawed losses or incorrect resetting of the hidden states; (b) ineffective or misleading hyperparameters, e.g.~dropout probability meaning keep probability or setting the dropout not actually enabling dropout; (c) inefficiencies, e.g.~negative sampling after score computations. The bugs of each reimplementation are discussed in Section~\ref{sec:quant}.
\section{Quantitative comparison}\label{sec:quant}

Section~\ref{sec:compare} showed that none of the reimplementations are feature complete, and most of them have bugs. However, it is possible that the missing features are not that important or even the bugs might have only a small impact on the performance of the algorithm. In this section we examine how these differences affect the performance of the algorithm (RQ4). This section provides quantitative comparison between the official and third-party implementations.

\subsection{Datasets}

We use five publicly available, real-life session-based datasets.
\begin{itemize}
\item The \textbf{Yoochoose}\footnote{\url{https://2015.recsyschallenge.com}}\footnote{\url{https://www.kaggle.com/datasets/chadgostopp/recsys-challenge-2015}} dataset contains user sessions from an unnamed e-commerce site. The data is already split into proper sessions and each session consists of one or more click events and might have purchase event(s) associated with it. As we are focusing on the next item prediction task within sessions, we only use the click events. The dataset was originally released for RecSys Challenge 2015 and has been used for evaluating session-based recommenders since then, e.g.~by~\cite{hidasi2015session,hidasi2018recurrent,de2021transformers4rec}.
\item The \textbf{Rees46}\footnote{\url{https://www.kaggle.com/datasets/mkechinov/ecommerce-behavior-data-from-multi-category-store}}\footnote{\url{https://rees46.com/en/open-cdp}} dataset contains 8 months of user behavior data -- view, cart and purchase events -- from a multicategory e-commerce website between October 2019 and April 2020. Similarly to~\cite{de2021transformers4rec} we only used some of the 8 months, but whereas they used only one month, we use two\footnote{Only these two months are available directly from the dataset's page, the rest of the data is available through an external site.}. This is similar how real-life recommenders are usually trained only on the most recent $0.5$--$2$ months to avoid concept drift, if the traffic is large enough. We only use view events for next item prediction. The dataset does come with precomputed sessions, but it is unclear how user histories are split into sessions. E.g.~a few sessions have events from multiple users, and the time gap between subsequent events within a session can be arbitrarily long. Therefore, sessions were recomputed from the user histories using the standard 1 hour session gap threshold (i.e.~if the time gap between two subsequent events within a user's history is longer than 1 hour, the later event is the start of a new session).
\item The \textbf{Coveo}\footnote{\url{https://github.com/coveooss/shopper-intent-prediction-nature-2020}}~\cite{requena2020shopper} dataset was released for the purpose of shopper intent prediction from clickstream data as part of the SIGIR Ecom 2021 Challenge. The dataset consists of user sessions using 30 minutes as the session gap threshold. The data was collected by Coveo from one of it's partner's e-commerce site. There are five types of events: add, click, detail, purchase and remove. Detail events correspond to the user visiting the detail page of an item (similarly to view events in e.g.~Yoochoose). We use only these detail events to build the sessions.
\item The \textbf{RetailRocket} dataset\footnote{\url{https://www.kaggle.com/datasets/retailrocket/ecommerce-dataset}} consists of view, add to cart, and transaction events collected over 4.5 months. Similarly to previous datasets, we only keep the view events. The dataset was published to motivate research in the field of recommender systems and was used by e.g.~\cite{xu2019graph,gupta2019niser,luo2020collaborative,xin2020self}. 
\item The \textbf{Diginetica}\footnote{\url{https://competitions.codalab.org/competitions/11161\#learn\_the\_details-data2}} dataset consists of user sessions collected over six months. The sessions were extracted from the search engine logs of an e-commerce site, and contain item page views that were preceded by search queries. The dataset doesn't have exact timestamps, but each event has a property denoting the elapsed time since its session's first query. The day of the first queries is also known. The dataset was released for the CIKM Cup 2016 Track 2: Personalized E-Commerce Search Challenge. Papers using the dataset include~\cite{wu2019session,xu2019graph,wang2020global}.
\end{itemize}

Data preprocessing is similar to~\cite{hidasi2015session,hidasi2018recurrent} with an additional step to approximate real-life use-cases even better:
\begin{enumerate}
  \item Only view~/~click~/~detail events are used (depending on the dataset), since our focus is on the next item prediction task as it was defined in~\cite{hidasi2015session,hidasi2018recurrent}. These events are associated with the user accessing an item detail page either through recommendations or organically, and together they describe item sequences or sessions.
  \item User histories are split into sessions using 1 hour as the session gap threshold for Rees46 and RetailRocket. Precomputed sessions are used for Yoochoose, Coveo and Diginetica.
  \item Any data except the session ID, the item ID and the timestamp is ignored. For Diginetica a (virtual) timestamp is computed from the day of the first query and the elapsed time.
  \item Subsequent repeating items in the sessions are filtered. I.e.~if the user visits the same item multiple times in succession, only the first occurrence is kept. E.g.~$(i,i,j)$ is reduced to $(i,j)$, but $(i,j,i)$ is not modified. This is an extra preprocessing step compared to \cite{hidasi2015session,hidasi2018recurrent}. We argue that this aligns better with the goal of next item recommendation for both training and evaluation. Repeating events can occur either by the user reloading/reopening a page or via logging errors. Neither of these are informative for recommenders as recommending the same item the user is currently viewing is never useful.
  \item The dataset is iteratively filtered for sessions shorter than 2 and items occurring less than 5 times until there is no change in the dataset. I.e.~after this filtering, every session consists of at least 2 items and each item is present in the dataset at least 5 times. Sessions of length 1 are useless for both training and evaluation. A very low minimal item support threshold is enforced, due to the collaborative filtering nature of GRU4Rec and to be consistent with the original papers.
  \item Train/test splits are time based and very similar to~\cite{hidasi2015session,hidasi2018recurrent}. The split time for Yoochoose, Rees46 and Coveo is 1 day before the time of the dataset's last event, and 7 days for RetailRocket and Diginetica. The latter two datasets don't have enough events to use only one day for evaluation. The test set consists of sessions that started after the split time. The train set consists of events that happened before the split time, i.e.~sessions extending over are cut off at the split time.
\end{enumerate}

\begin{table*}
  \caption{Basic statistics of the datasets (train/test split)}
  \label{tab:data_stat}
  \small
  \begin{tabular}{l|rrrr|rrrr|r}
    \toprule
    \multirow{2}{*}{Dataset} & \multicolumn{4}{c|}{Training set} & \multicolumn{4}{c|}{Test set} & \multirow{2}{*}{\#Items}\\
    & \multicolumn{1}{c}{\#Events} & \multicolumn{1}{c}{\#Sessions} & \multicolumn{1}{c}{\#Days} & \multicolumn{1}{c|}{Events/session} & \multicolumn{1}{c}{\#Events} & \multicolumn{1}{c}{\#Sessions} & \multicolumn{1}{c}{\#Days} & \multicolumn{1}{c|}{Events/session} &\\
    \midrule
    Yoochoose & 29,107,309 & 7,597,703 & 181 & 3.83 & 71,849 & 15,854 & 1 & 4.53 & 34,858 \\
    Rees46 & 67,575,203 & 10,190,006 & 60 & 6.63 & 1,054,210 & 166,841 & 1 & 6.32 & 172,756 \\
    Coveo & 1,411,113 & 165,673 & 17 & 8.52 & 52,501 & 7,748 & 1 & 6.78 & 10,868 \\
    RetailRocket & 750,832 & 196,234 & 131 & 3.83 & 29,148 & 8,036 & 7 & 3.63 & 36,824 \\
    Diginetica & 833,113 & 177,266 & 146 & 4.70 & 70,164 & 15,040 & 7 & 4.67 & 40,351 \\
    \bottomrule
  \end{tabular}
\end{table*}

Table~\ref{tab:data_stat} summarizes the basic statistics of the dataset. Note that after the train/test split, a few sessions of the train set might have a length of 1 due to the strict cut-off. These are filtered from the train set. Test sets might contain items that are not in the train sets. Since GRU4Rec is not able to handle item cold-start, we skip these during evaluation. Preprocessing scripts and the code of our experiments are available on Github\footnote{\url{https://github.com/hidasib/gru4rec_third_party_comparison}}.

\subsection{Evaluation setup}

We use the \emph{next item prediction} task as defined in~\cite{hidasi2015session,hidasi2018recurrent}. Apart from the first event of each session, a recommendation list is generated for every test event. This list is based solely on the events that happened before the test event in the same session. The goal of the model is to rank the item of the test event as high as possible among the recommendations.

We measure \emph{recall@N} and \emph{MRR@N}. Recall@N is the ratio of test items present in their corresponding recommendation lists of length $N$ and the number of test items. Since in this task a separate recommendation list is generated for every test item, recall@N is the same as hitrate@N. MRR@N (mean reciprocal rank at $N$) is the mean of the reciprocal ranks of the test items in their corresponding recommendation lists of length $N$. If the test item is not in the recommendation list, its reciprocal rank is considered to be 0. The metrics were measured at $N=\{1,5,10,20\}$. Similar trends can be observed for different values of $N$, thus only results at $N=20$ are included in the paper due to space limitations. However, every result is available at the website\footnote{\url{https://hidasib.github.io/gru4rec_third_party_comparison}} of the study.

Section~\ref{sec:compare} revealed that the reimplementations lack some features of the original, and also suffer from implementation bugs. Aside from the direct out-of-the-box comparison, we are also interested in what is achievable with the reimplementation by researchers of various thoroughness. Therefore, we created improved versions of the third-party implementations as if an experienced user would have reviewed the code and fixed a portion of the bugs:
\begin{enumerate}
  \item \textbf{Out-of-the-box:} Only errors preventing the execution of the code are fixed.
  \item \textbf{Inference fix:} Only errors of the inferencing code are fixed. (E.g.~inappropriate resetting of the hidden state.)
  \item \textbf{Minor fix:} Easily noticeable and fixable bugs of the training are fixed. Hard-coded parameters are made settable.
  \item \textbf{Major fix:} Obscure and/or hard to fix errors that still don't require architectural change or rewriting a significant portion of the code are fixed. Note that this version is still not fully fixed.
\end{enumerate}
Note that (a) fixes are incremental, i.e.~higher level fixes contain lower level ones; (b) not all levels are relevant for every third-party implementation; (c) missing GRU4Rec features are not added to the code in either version. 

The effect of bugs and the lack of features are examined separately. Therefore, the official GRU4Rec is trained both with all of its features and feature sets matching the capabilities of the reimplementations (using their highest level fix).

Hyperparameters are optimized on a separate training/validation split. This is created from the full training set using the same process as the train/test split. Optimization is performed with the feature complete official version on all five of the datasets separately, using Optuna\footnote{https://optuna.org/} over 200 iterations. The models are then retrained on the full training set using the optimal hyperparameters or the supported subset corresponding to the capabilities of reimplementations with missing parameters. Results presented in the paper are measured on the test set. Parameter search spaces, optimal parameters, and the parametrization of every experiment are available on the website of the study. 

\subsection{Results}
Every third-party implementation is compared to the full and matching features versions of the official implementation on all five datasets using both cross-entropy and the BPR-max losses, if the reimplementation supports both.

\subsubsection{GRU4REC-pytorch}\label{sssec:gru4rec_pytorch}

This reimplementation suffers from several hard to spot and/or fix bugs, and it is the only one with incorrect inference code. The following versions were used for our experiments:

\textbf{Out-of-the-box:} Running the code on GPU required moving mean computation of a variable to the correct device.

\textbf{Inference fix:} The evaluation code now resets the hidden state when the corresponding session ends.

\textbf{Major fix:} (a) Fixed the order of sampling and applying softmax transformation, as it was in the reverse order resulting in small gradients and slow convergence. (b) Softmax transformation is now only applied once (was twice). (c) Hidden states are now reset correctly during training. The mask governing the resets was only recalculated when a session ended, resulting in false resets. (d) BPR-max loss is fixed to use the correct equation, but the missing score regularization was not added to algorithm. (e) Both dropout parameters now work as expected. Dropout on the final GRU layer and embedding dropout in separate embedding mode was originally not applied.

Sampling is performed after all item scores are computed, which slows down training. This bug is rooted so deep in the code that we did not fix it.

As shown in Table~\ref{tab:res_gru4rec-pytorch}, the out-of-the-box performance is very bad ($75\%$--$99\%$ lower than the original), even if inferencing is fixed ($51\%$--$96\%$ lower) due mostly to errors in the implementation of both loss functions. The rest of the bugs have a smaller, but still noticeable influence on the results. Missing features also significantly limit the implementation as it can be seen by comparing the matching and all features versions of the official implementation.

\begin{table*}
  \caption{Quantitative comparison of GRU4REC-pytorch to the official implementation using recall@20 and MRR@20. Relative difference compared to the official version is shown in parentheses.}
  \label{tab:res_gru4rec-pytorch}
  \scriptsize
  \begin{tabular}{l|ll|ll|ll|ll|ll}
    \toprule
    \multirow{2}{*}{Dataset}&\multicolumn{4}{c|}{Official GRU4Rec}&\multicolumn{6}{c}{GRU4REC-pytorch}\\
    &\multicolumn{2}{c}{All features}&\multicolumn{2}{c|}{Matching features}&\multicolumn{2}{c}{Out-of-the-box}&\multicolumn{2}{c}{Inference fix}&\multicolumn{2}{c}{Major fix}\\
    \midrule
    \multicolumn{11}{c}{Cross-entropy loss}\\
    \midrule
    &\multicolumn{1}{c}{Recall}&\multicolumn{1}{c|}{MRR}&\multicolumn{1}{c}{Recall}&\multicolumn{1}{c|}{MRR}&\multicolumn{1}{c}{Recall}&\multicolumn{1}{c|}{MRR}&\multicolumn{1}{c}{Recall}&\multicolumn{1}{c|}{MRR}&\multicolumn{1}{c}{Recall}&\multicolumn{1}{c}{MRR}\\
    \midrule
    Yoochoose & 0.6804 & 0.3002 & 0.4583 (-33\%) & 0.1523 (-49\%) & 0.1169 (-83\%) & 0.1006 (-66\%)& 0.1212 (-82\%) & 0.1037 (-65\%) & 0.4271 (-37\%) & 0.1227 (-59\%) \\
    Rees46 & 0.5291 & 0.2003 & 0.3059 (-42\%) & 0.0828 (-59\%) & 0.1028 (-81\%) & 0.0713 (-64\%) & 0.1093 (-79\%) & 0.0755 (-62\%) & 0.1994 (-62\%) & 0.0371 (-81\%) \\
    Coveo & 0.2947 & 0.0960 & 0.2061 (-30\%) & 0.0591 (-38\%) & 0.0628 (-79\%) & 0.0386 (-60\%) & 0.0663 (-78\%) & 0.0397 (-59\%) & 0.1806 (-39\%) & 0.0497 (-48\%) \\
    RetailRocket & 0.4953 & 0.2012 & 0.3378 (-32\%) & 0.1121 (-44\%) & 0.0616 (-88\%) & 0.0517 (-74\%) & 0.0642 (-87\%) & 0.0545 (-73\%) & 0.2812 (-43\%) & 0.0986 (-51\%) \\
    Diginetica & 0.4874 & 0.1440 & 0.2973 (-39\%) & 0.0749 (-48\%) & 0.0457 (-91\%) & 0.0329 (-77\%) & 0.0503 (-90\%) & 0.0366 (-75\%) & 0.2862 (-41\%) & 0.0736 (-49\%) \\
    \midrule
    \multicolumn{11}{c}{BPR-max loss}\\
    \midrule
    &\multicolumn{1}{c}{Recall}&\multicolumn{1}{c|}{MRR}&\multicolumn{1}{c}{Recall}&\multicolumn{1}{c|}{MRR}&\multicolumn{1}{c}{Recall}&\multicolumn{1}{c|}{MRR}&\multicolumn{1}{c}{Recall}&\multicolumn{1}{c|}{MRR}&\multicolumn{1}{c}{Recall}&\multicolumn{1}{c}{MRR}\\
    \midrule
    Yoochoose & 0.6799 & 0.2931 & 0.4655 (-32\%) & 0.1760 (-40\%) & 0.0087 (-99\%) & 0.0011 (-99\%) & 0.1066 (-84\%) & 0.0118 (-96\%) & 0.1923 (-72\%) & 0.0391 (-87\%) \\
    Rees46 & 0.5206 & 0.1913 & 0.2615 (-50\%) & 0.0691 (-64\%) & 0.1316 (-75\%) & 0.0225 (-88\%) & 0.1418 (-73\%) & 0.0232 (-88\%) & 0.1764 (-66\%) & 0.0355 (-81\%) \\
    Coveo & 0.3123 & 0.0994 & 0.2216 (-29\%) & 0.0643 (-35\%) & 0.1141 (-63\%) & 0.0279 (-72\%) & 0.1510 (-52\%) & 0.0368 (-63\%) & 0.1582 (-49\%) & 0.0410 (-59\%) \\
    RetailRocket & 0.5187 & 0.2131 & 0.3142 (-39\%) & 0.1067 (-50\%) & 0.0794 (-85\%) & 0.0227 (-89\%) & 0.2528 (-51\%) & 0.0848 (-60\%) & 0.2554 (-51\%) & 0.0916 (-57\%) \\
    Diginetica & 0.4995 & 0.1500 & 0.3141 (-37\%) & 0.0851 (-43\%) & 0.0070 (-99\%) & 0.0015 (-99\%) & 0.2367 (-53\%) & 0.0608 (-59\%) & 0.2616 (-48\%) & 0.0692 (-54\%) \\
  \bottomrule
\end{tabular}
\end{table*}

\subsubsection{Torch-GRU4Rec}

The only error in \emph{Torch-GRU4Rec} is that sampling is performed after the scores are computed, which increases training times. This bug is rooted so deep in the code that we did not fix it. Therefore, this reimplementation only has an out-of-the-box version:

\textbf{Out-of-the-box:} A single typo needed to be fixed, so the code is able to run.

The small differences (e.g.~missing momentum parameter) add up to a noticeable difference when Torch-GRU4Rec is compared to the matching features version. However, the performance of this reimplementation suffers the most due the missing constrained embedding mode (up to $19\%$ lower performance) as shown in Table~\ref{tab:res_torch-gru4rec}.

\begin{table*}
  \caption{Quantitative comparison of Torch-GRU4Rec to the official implementation using recall@20 and MRR@20. Relative difference compared to the official version is shown in parentheses.}
  \label{tab:res_torch-gru4rec}
  \small
  \begin{tabular}{l|ll|ll|ll}
    \toprule
    \multirow{2}{*}{Dataset}&\multicolumn{4}{c|}{Official GRU4Rec}&\multicolumn{2}{c}{Torch-GRU4Rec}\\
    &\multicolumn{2}{c}{All features}&\multicolumn{2}{c|}{Matching features}&\multicolumn{2}{c}{Out-of-the-box}\\
    \midrule
    \multicolumn{7}{c}{Cross-entropy loss}\\
    \midrule
    &\multicolumn{1}{c}{Recall}&\multicolumn{1}{c|}{MRR}&\multicolumn{1}{c}{Recall}&\multicolumn{1}{c|}{MRR}&\multicolumn{1}{c}{Recall}&\multicolumn{1}{c}{MRR}\\
    \midrule
    Yoochoose & 0.6804 & 0.3002 & 0.6690 (-2\%) & 0.2882 (-4\%) & 0.6671 (-2\%) & 0.2847 (-5\%) \\
    Rees46 & 0.5291 & 0.2003 & 0.4716 (-11\%) & 0.1624 (-19\%) & 0.4688 (-11\%) & 0.1606 (-20\%) \\
    Coveo & 0.2947 & 0.0960 & 0.2848 (-3\%) & 0.0928 (-3\%) & 0.2835 (-4\%) & 0.0897 (-7\%) \\
    RetailRocket & 0.4953 & 0.2012 & 0.4249 (-14\%) & 0.1635 (-19\%) & 0.3834 (-23\%) & 0.1464 (-27\%) \\
    Diginetica & 0.4874 & 0.1440 & 0.4255 (-13\%) & 0.1246 (-13\%) & 0.4245 (-13\%) & 0.1229 (-15\%) \\
    \midrule
    \multicolumn{7}{c}{BPR-max loss}\\
    \midrule
    &\multicolumn{1}{c}{Recall}&\multicolumn{1}{c|}{MRR}&\multicolumn{1}{c}{Recall}&\multicolumn{1}{c|}{MRR}&\multicolumn{1}{c}{Recall}&\multicolumn{1}{c}{MRR}\\
    \midrule
    Yoochoose & 0.6799 & 0.2931 & 0.6769 (-0\%) & 0.2919 (-0\%) & 0.6711 (-1\%) & 0.2907 (-1\%) \\
    Rees46 & 0.5206 & 0.1913 & 0.5112 (-2\%) & 0.1839 (-4\%) & 0.5081 (-2\%) & 0.1820 (-5\%) \\
    Coveo & 0.3123 & 0.0994 & 0.2995 (-4\%) & 0.0962 (-3\%) & 0.2960 (-5\%) & 0.0944 (-5\%) \\
    RetailRocket & 0.5187 & 0.2131 & 0.4639 (-11\%) & 0.1773 (-17\%) & 0.4201 (-19\%) & 0.1633 (-23\%) \\
    Diginetica & 0.4995 & 0.1500 & 0.4597 (-8\%) & 0.1383 (-8\%) & 0.4550 (-9\%) & 0.1372 (-9\%) \\
  \bottomrule
\end{tabular}
\end{table*}

\subsubsection{GRU4Rec\_Tensorflow}

The following versions were created for GRU4Rec\_Tensorflow:

\textbf{Out-of-the-box:} The code is able to run, however it is not prepared for unknown test items. The dropout parameter was fixed to regain its original meaning as it was used as keep probability instead of drop probability.

\textbf{Minor fix:} (a) Lowered the large initial accumulator value close to zero as it prevented sufficient convergence. (b) The exponential learning rate decay is discarded as it is not part of the original. (c) Now the hard-coded \texttt{batch\_size} parameter can be changed.

Out-of-the-box performance is poor ($43\%$--$94\%$ lower compared to the original) due to the error preventing proper convergence. The fixed version is still significantly worse ($32\%$--$67\%$ lower), but most of the accuracy loss results from the missing features and not the bugs. See Table~\ref{tab:res_gru4rec_tensorflow} for details.

\begin{table*}
  \caption{Quantitative comparison of GRU4Rec\_Tensorflow to the official implementation using recall@20 and MRR@20. Relative difference compared to the official version is shown in parentheses.}
  \label{tab:res_gru4rec_tensorflow}
  \small
  \begin{tabular}{l|ll|ll|ll|ll}
    \toprule
    \multirow{2}{*}{Dataset}&\multicolumn{4}{c|}{Official GRU4Rec}&\multicolumn{4}{c}{GRU4Rec\_Tensorflow}\\
    &\multicolumn{2}{c}{All features}&\multicolumn{2}{c|}{Matching features}&\multicolumn{2}{c}{Out-of-the-box}&\multicolumn{2}{c}{Minor fix}\\
    \midrule
    \multicolumn{9}{c}{Cross-entropy loss}\\
    \midrule
    &\multicolumn{1}{c}{Recall}&\multicolumn{1}{c|}{MRR}&\multicolumn{1}{c}{Recall}&\multicolumn{1}{c|}{MRR}&\multicolumn{1}{c}{Recall}&\multicolumn{1}{c|}{MRR}&\multicolumn{1}{c}{Recall}&\multicolumn{1}{c}{MRR}\\
    \midrule
    Yoochoose & 0.6804 & 0.3002 & 0.4555 (-33\%) & 0.1495 (-50\%) & 0.3857 (-43\%) & 0.1359 (-55\%) & 0.4606 (-32\%) & 0.1482 (-51\%) \\
    Rees46 & 0.5291 & 0.2003 & 0.2940 (-44\%) & 0.0765 (-62\%) & 0.2066 (-61\%) & 0.0594 (-70\%) & 0.2698 (-49\%) & 0.0667 (-67\%) \\
    Coveo & 0.2947 & 0.0960 & 0.2001 (-32\%) & 0.0576 (-40\%) & 0.0375 (-87\%) & 0.0075 (-92\%) & 0.1951 (-34\%) & 0.0549 (-43\%) \\
    RetailRocket & 0.4953 & 0.2012 & 0.3309 (-33\%) & 0.1092 (-46\%) & 0.0439 (-91\%) & 0.0212 (-89\%) & 0.3388 (-32\%) & 0.1179 (-41\%) \\
    Diginetica & 0.4874 & 0.1440 & 0.2976 (-39\%) & 0.0761 (-47\%) & 0.0284 (-94\%) & 0.0088 (-94\%) & 0.3014 (-38\%) & 0.0805 (-44\%) \\
  \bottomrule
\end{tabular}
\end{table*}

\subsubsection{Keras GRU4Rec}\label{sssec:keras_res}

\emph{Keras GRU4Rec} is probably the most simplistic of the reimplementations, thus it doesn't have that many serious bugs in its code.

\textbf{Out-of-the box:} The implementation as is.

\textbf{Minor fix:} (a) Hard-coded parameters (\textit{hidden\_size}, \textit{dropout\_p\_hidden}, \textit{learning\_rate}) now can be set. (b) The default optimizer is changed to Adagrad.

\textbf{Major fix:} (a) Fixed incorrect resetting of hidden states (the same error that GRU4REC-pytorch has). (b) Epochs don't end now when the number of remaining sessions is not enough to fully fill the mini-batch.

Table~\ref{tab:res_kerasgru4rec} summarizes the outcome of the experiments. The limited algorithmic feature set results in up to $15\%$ lower performance. The lack of BPR-max loss might further widen the gap between this version and the best performing model for certain datasets. Hard-coded parameters increase accuracy loss by $6\%$--$36\%$. Fixing the bugs allows the model to reach the level of official implementation with matching features on some of the datasets. Note that this version of the official implementation still uses negative item sampling during training. While it is possible to modify the official version to abandon sampling it is ultimately pointless as poor scaling would make it unusable in practice.

\begin{table*}
  \caption{Quantitative comparison of KerasGRU4Rec to the official implementation using recall@20 and MRR@20. Relative difference compared to the official version is shown in parentheses.}
  \label{tab:res_kerasgru4rec}
  \scriptsize
  \begin{tabular}{l|ll|ll|ll|ll|ll}
    \toprule
    \multirow{2}{*}{Dataset}&\multicolumn{4}{c|}{Official GRU4Rec}&\multicolumn{6}{c}{KerasGRU4Rec}\\
    &\multicolumn{2}{c}{All features}&\multicolumn{2}{c|}{Matching features}&\multicolumn{2}{c}{Out-of-the-box}&\multicolumn{2}{c}{Minor fix}&\multicolumn{2}{c}{Major fix}\\
    \midrule
    \multicolumn{11}{c}{Cross-entropy loss}\\
    \midrule
    &\multicolumn{1}{c}{Recall}&\multicolumn{1}{c|}{MRR}&\multicolumn{1}{c}{Recall}&\multicolumn{1}{c|}{MRR}&\multicolumn{1}{c}{Recall}&\multicolumn{1}{c|}{MRR}&\multicolumn{1}{c}{Recall}&\multicolumn{1}{c|}{MRR}&\multicolumn{1}{c}{Recall}&\multicolumn{1}{c}{MRR}\\
    \midrule
    Yoochoose & 0.6804 & 0.3002 & 0.6722 (-1\%) & 0.2919 (-3\%) & 0.6392 (-6\%) & 0.2666 (-11\%) & 0.6784 (-0\%) & 0.3010 (+0\%) & 0.6768 (-1\%) & 0.3013 (+0\%) \\
    Rees46 & 0.5291 & 0.2003 & 0.4860 (-8\%) & 0.1740 (-13\%) & 0.4608 (-13\%) & 0.1574 (-21\%) & 0.4979 (-6\%) & 0.1868 (-7\%) & 0.4978 (-6\%) & 0.1866 (-7\%) \\
    Coveo & 0.2947 & 0.0960 & 0.2852 (-3\%) & 0.0917 (-4\%) & 0.2739 (-7\%) & 0.0871 (-9\%) & 0.2822 (-4\%) & 0.0925 (-4\%) & 0.2826 (-4\%) & 0.0918 (-4\%) \\
    RetailRocket & 0.4953 & 0.2012 & 0.4239 (-14\%) & 0.1705 (-15\%) & 0.3151 (-36\%) & 0.1250 (-38\%) & 0.3938 (-20\%) & 0.1642 (-18\%) & 0.3957 (-20\%) & 0.1647 (-18\%) \\
    Diginetica & 0.4874 & 0.1440 & 0.4304 (-12\%) & 0.1268 (-12\%) & 0.3922 (-20\%) & 0.1133 (-21\%) & 0.3853 (-21\%) & 0.1146 (-20\%) & 0.3869 (-21\%) & 0.1141 (-21\%) \\
  \bottomrule
\end{tabular}
\end{table*}

\subsubsection{GRU4Rec in Recpack}

The following versions were used for testing Recpack GRU4Rec:

\textbf{Out-of-the-box:} Running the code on GPU required fixes in the evaluation pipeline and the application of dropout.

\textbf{Minor fix:} (a) \texttt{bpreg} is no longer hard-coded. (b) Fixed BPR-max so $log()$ can no longer have $0$ as its argument. (c) Switched weight matrix initialization to Glorot uniform~\cite{glorot2010understanding} to match the original.

\textbf{Major fix:} Fixed dropout to use separate parameters for embeddings and hidden states.

Similarly to GRU4REC-pytorch, this version also suffers from computing the scores before sampling. This is rooted too deep in the code to be fixed.

The missing features already amount for up to $23\%$ accuracy loss, which is increased up to $\sim51\%$ due to bugs in the code. Note that the matching features version of the official algorithm here uses negative sampling (see \ref{sssec:keras_res}).

\begin{table*}
  \caption{Quantitative comparison of Recpack GRU4Rec to the official implementation using recall@20 and MRR@20. Relative difference compared to the official version is shown in parentheses.}
  \label{tab:res_recpackgru4rec}
  \scriptsize
  \begin{tabular}{l|ll|ll|ll|ll|ll}
    \toprule
    \multirow{2}{*}{Dataset}&\multicolumn{4}{c|}{Official GRU4Rec}&\multicolumn{6}{c}{Recpack GRU4Rec}\\
    &\multicolumn{2}{c}{All features}&\multicolumn{2}{c|}{Matching features}&\multicolumn{2}{c}{Out-of-the-box}&\multicolumn{2}{c}{Minor fix}&\multicolumn{2}{c}{Major fix}\\
    \midrule
    \multicolumn{11}{c}{Cross-entropy loss}\\
    \midrule
    &\multicolumn{1}{c}{Recall}&\multicolumn{1}{c|}{MRR}&\multicolumn{1}{c}{Recall}&\multicolumn{1}{c|}{MRR}&\multicolumn{1}{c}{Recall}&\multicolumn{1}{c|}{MRR}&\multicolumn{1}{c}{Recall}&\multicolumn{1}{c|}{MRR}&\multicolumn{1}{c}{Recall}&\multicolumn{1}{c}{MRR}\\
    \midrule
    Yoochoose & 0.6804 & 0.3002 & 0.6713 (-1\%) & 0.2877 (-4\%) & 0.6214 (-9\%) & 0.2627 (-12\%) & 0.6256 (-8\%) & 0.2653 (-12\%) & 0.6382 (-6\%) & 0.2755 (-8\%) \\
    Rees46 & 0.5291 & 0.2003 & 0.4812 (-9\%) & 0.1655 (-17\%) & 0.4412 (-17\%) & 0.1490 (-26\%) & 0.4414 (-17\%) & 0.1502 (-25\%) & 0.4386 (-17\%) & 0.1496 (-25\%) \\
    Coveo & 0.2947 & 0.0960 & 0.2829 (-4\%) & 0.0912 (-5\%) & 0.2335 (-21\%) & 0.0752 (-22\%) & 0.2323 (-21\%) & 0.0739 (-23\%) & 0.2309 (-22\%) & 0.0722 (-25\%) \\
    RetailRocket & 0.4953 & 0.2012 & 0.4235 (-14\%) & 0.1637 (-19\%) & 0.2442 (-51\%) & 0.1029 (-49\%) & 0.2210 (-55\%) & 0.0927 (-54\%) & 0.2866 (-42\%) & 0.1157 (-42\%) \\
    Diginetica & 0.4874 & 0.1440 & 0.4263 (-13\%) & 0.1248 (-13\%) & 0.3279 (-33\%) & 0.0937 (-35\%) & 0.3219 (-34\%) & 0.0927 (-36\%) & 0.3296 (-32\%) & 0.0944 (-34\%) \\
    \midrule
    \multicolumn{11}{c}{BPR-max loss}\\
    \midrule
    &\multicolumn{1}{c}{Recall}&\multicolumn{1}{c|}{MRR}&\multicolumn{1}{c}{Recall}&\multicolumn{1}{c|}{MRR}&\multicolumn{1}{c}{Recall}&\multicolumn{1}{c|}{MRR}&\multicolumn{1}{c}{Recall}&\multicolumn{1}{c|}{MRR}&\multicolumn{1}{c}{Recall}&\multicolumn{1}{c}{MRR}\\
    \midrule
    Yoochoose & 0.6799 & 0.2931 & 0.6749 (-1\%) & 0.2919 (-0\%) & 0.6430 (-5\%) & 0.2619 (-11\%) & 0.6464 (-5\%) & 0.2697 (-8\%) & 0.6433 (-5\%) & 0.2691 (-8\%) \\
    Rees46 & 0.5206 & 0.1913 & 0.5010 (-4\%) & 0.1743 (-9\%) & 0.4624 (-11\%) & 0.1536 (-20\%) & 0.4624 (-11\%) & 0.1544 (-19\%) & 0.4666 (-10\%) & 0.1562 (-18\%) \\
    Coveo & 0.3123 & 0.0994 & 0.3003 (-4\%) & 0.0951 (-4\%) & 0.2091 (-33\%) & 0.0612 (-38\%) & 0.1953 (-37\%) & 0.0573 (-42\%) & 0.1536 (-51\%) & 0.0435 (-56\%) \\
    RetailRocket & 0.5187 & 0.2131 & 0.4455 (-14\%) & 0.1647 (-23\%) & 0.3226 (-38\%) & 0.1157 (-46\%) & 0.3054 (-41\%) & 0.1092 (-49\%) & 0.3550 (-32\%) & 0.1245 (-42\%) \\
    Diginetica & 0.4995 & 0.1500 & 0.4586 (-8\%) & 0.1379 (-8\%) & 0.3066 (-39\%) & 0.0913 (-39\%) & 0.3744 (-25\%) & 0.1046 (-30\%) & 0.3756 (-25\%) & 0.1040 (-31\%) \\
  \bottomrule
\end{tabular}
\end{table*}

\subsection{Training times \& scalability}\label{ssec:scaling}

Scalability is really important for recommender algorithms as most real-life datasets are large, e.g.~Rees46 can be considered to be average sized. Table~\ref{tab:training_times} shows the average time of a single epoch of the out-of-the-box version of every implementation on all datasets (RQ5), measured on an nVidia A30. Note that the value of the hyperparameters and the number of events in the training data can significantly affect training times. Therefore, the baseline on each dataset--parameterization pair can be significantly different.

As mentioned in the previous section, KerasGRU4Rec and the Recpack reimplementations are scaling poorly, thus training them might take more than $300+$ times longer than the original. On Rees46, these two algorithms are unusable, because a single epoch takes $0.5\dots2$ days. The rest of the third-party implementations are usually $2-5$ times slower that can be attributed to the different deep learning frameworks they utilize. However, scalability and optimization of these reimplementations is also far from perfect as they mildly break down on the Rees46 dataset.

\begin{table*}
  \caption{Training time (in seconds) of one epoch of the official and third party GRU4Rec implementations. The value in parentheses show how much longer the training of reimplementation takes over the official version.}
  \label{tab:training_times}
  \small
  \begin{tabular}{l|r|r|r|r|r|r}
    \toprule
    Dataset & GRU4Rec & GRU4Rec-pytorch & Torch-GRU4Rec & GRU4Rec\_TFlow & KerasGRU4Rec & \multicolumn{1}{c}{Recpack}\\
    \midrule
    \multicolumn{7}{c}{Cross-entropy loss}\\
    \midrule
    Yoochoose & 452 & 1,948 (x4.31) & 2,082 (x4.61) & 982 (x2.17) & 13,740 (x30.4) & 13,458 (x29.77) \\
    Rees46 & 367 & 7,618 (x20.76) & 7,193 (x19.6) & 2,382 (x6.49) & 123,266 (x335.87) & 49,262 (x134.23) \\
    Coveo & 37 & 85 (x2.3) & 91 (x2.46) & 55 (x1.49) & 568 (x15.35) & 663 (x17.92) \\
    RetailRocket & 2.8 & 10.1 (x3.61) & 10.7 (x3.82) & 23 (x8.21) & 276.3 (x98.68) & 122.3 (x43.68) \\
    Diginetica & 4.5 & 17.7 (x3.93) & 17.7 (x3.93) & 26.3 (x5.84) & 359.3 (x79.84) & 115.2 (x25.6) \\
    \midrule
    \multicolumn{7}{c}{BPR-max loss}\\
    \midrule
    Yoochoose & 488 & 1,855 (x3.8) & 2,164 (x4.43) & N/A & N/A & 21,270 (x43.59) \\
    Rees46 & 1,957 & 29,528 (x15.09) & 30,690 (x15.68) & N/A & N/A & 179,835 (x91.89) \\
    Coveo & 12 & 23 (x1.92) & 32 (x2.67) & N/A & N/A & 771 (x64.25) \\
    RetailRocket & 6.9 & 20.5 (x2.97) & 26.4 (x3.83) & N/A & N/A & 450.2 (x65.25) \\
    Diginetica & 8 & 23.2 (x2.9) & 36.9 (x4.61) & N/A & N/A & 352.2 (x44.03) \\
    \bottomrule
  \end{tabular}
  \end{table*}

\section{Discussion}\label{sec:discuss}
The comparison of the six third-party reimplementations with the original resulted in this final tally:
\begin{itemize}
    \item Microsoft Recommender's version is GRU4Rec in name only. It implements a different and flawed algorithm.
    \item The rest of the reimplementations lack at least one GRU4Rec feature that has significant impact on accuracy.
    \item Except for Torch-GRU4Rec, these unofficial implementations suffer from bugs, hard-coded parameters, and divergence from the original; which further deteriorates recommendation accuracy.
    \item KerasGRU4Rec and Recpack scale poorly due to the lack of negative item sampling during training, but all five suffer from some kind of scalability problem.
\end{itemize}

Even though we expected some of the third-party implementations being incorrect or lacking features, this outcome still comes as an unpleasant surprise. Only Torch-GRU4Rec can be deemed as a competent attempt: while it still lacks an important feature, it is mostly a correct and faithful implementation. GRU4Rec is not a complex algorithm, and its main features are highlighted in its papers. Its publicly available official code can be used as both a crutch during reimplementation and as a standard to validate against. Therefore, it is puzzling why the quality of these reimplementations is this low. Considering all of the above, we assume that the problem is not isolated or specific to GRU4Rec. Many algorithms might be affected by incorrect unofficial implementations. To somewhat alleviate the harm, we release feature complete reimplementations of GRU4Rec for both Tensorflow\footnote{\url{https://github.com/hidasib/GRU4Rec_Tensorflow_Official}} and PyTorch\footnote{\url{https://github.com/hidasib/GRU4Rec_PyTorch_Official}} that are validated against the official version.

As discussed in Section~\ref{sec:intro}, incorrect unofficial public implementations might have a devastating effect on multiple research projects, depending on how widespread they are. Unfortunately, most papers don't mention which implementation of the baselines were used during experimentation. Our findings suggest that this is important information, on par with optimal hyperparameters or the description of data preprocessing.

Our study raises some interesting questions for the research community to answer. If a research project showcases suboptimal baseline performance due to using an incorrect third-party implementation, who is responsible? Researchers are responsible for what they publish, but to what extent do they need to double-check whether the tools they use are correct? Does the implementation of every single baseline need to be validated rigorously against its original paper? Are the contributors of reimplementations responsible for the misleading results, even if code is provided as-is and without any guarantees? What should happen if a reimplementation is found incorrect? Is it feasible to correct experiments in a large number, sometimes several years old papers? Should the incorrect third-party implementations get fixed or withdrawn? What if these have been long abandoned by their original contributors?

Our advice to researchers is to either use the official implementation or validate the reimplementation they use against the official version, and clearly state in their paper which implementation they used. Official open source releases can also help reimplementation efforts, which is another reason for publishing code. Most importantly, contributors of third-party implementations should validate their work before releasing it to the public.
\section{Related work}\label{sec:related}
We were unable to find any studies closely relating to our work. Looking at the wider area~\cite{ferrari2019we,ferrari2020methodological} examined multiple top-n recommender algorithms presented at prestigious conferences, and found that only 7 out of 18 works could be reproduced with reasonable effort. Furthermore, 6 of those 7 could be outperformed by simple baselines. The major problems they highlight relate to using weak baselines, arbitrary evaluation setups and suboptimal hyperparameters. The latter was also examined in detail in~\cite{rendle2022revisiting} through the example of the now 15 years old iALS algorithm. iALS can be a much stronger baseline -- capable of beating more modern algorithms -- with the appropriate parameterization.

\cite{ferrari2021troubling} reassures the findings of~\cite{ferrari2019we,ferrari2020methodological} for a wider variety of algorithms and baselines. They argue that while the majority of recent papers utilize evaluation setups, datasets, metrics, etc.~of previous work, these are reused without questioning their validity, thus popular but flawed evaluation setups can spread in the community.

A recent work~\cite{petrov2022systematic} reinforces the notion that claimed state-of-the-art performance is often skewed or invalid through the example of BERT4Rec, as the results of the original paper can not be reproduced.

Standardization of evaluation through frameworks like \cite{10.1145/3523227.3551472,argyriou2020microsoft} can help reproducibility, if those frameworks are mostly free of the issues discussed above. However, defining the proper evaluation setups with datasets, and correctly implementing a wide variety of algorithms is not an easy task. Our findings show that algorithms reimplemented in benchmarking frameworks can suffer from serious flaws.
\section{Conclusion}\label{sec:conclusion}
When done correctly, reimplementing a popular algorithm can be beneficial for the research community due to increasing accessibility or contributing towards the standardization of evaluation. However, it may also pose immense risks to reproducibility if the unofficial implementation is incorrect or buggy. In this paper we described the results of a thorough comparison between six third-party and the official implementation of GRU4Rec, a seminal algorithm for session-based~/~sequential recommendations. One of the six unofficial reimplementations is GRU4Rec in its name only as it implements a different architecture, which is also conceptually flawed. The remaining five lack important features of the original that contribute to the good performance of the official version. Four of them also suffer from errors~/~bugs of varying severity that are detrimental to their performance. In the most extreme case, out-of-the-box performance is $99.63\%$ lower compared to the official version. The reimplementations are also significantly slower to train. Two of them scale very poorly, thus using them on larger datasets is highly impractical. We think that these findings are alarming. Potentially a significant number of research project outcomes have been falsified by these errors. But more importantly, we don't think that this problem is isolated and specific to GRU4Rec, which is not a too complicated algorithm with a publicly available official implementation. We urge the research community and especially those who reimplement algorithms -- either for their own use or as a public resource -- to validate reimplementations against the original work and the official implementation.

\begin{acks}
  The work leading to these results received funding from \grantsponsor{PIACIKFI}{National Research, Development and Innovation Office, Hungary}~~ under grant agreement number \grantnum{PIACIKFI}{2020-1.1.2-PIACI-KFI-2021-00289}.

  The authors would also like to thank Domonkos Tikk for his valuable support as the project leader of the above-mentioned R\&D grant.
\end{acks}

\bibliographystyle{ACM-Reference-Format}
\bibliography{references}

\end{document}